# From Classical to Wave-Mechanical Dynamics

Adriano Orefice[*], Raffaele Giovanelli, Domenico Ditto

*Università degli Studi di Milano - DISAA - Via Celoria, 2 - 20133 - Milano (Italy)*

ABSTRACT - The time-independent Schroedinger and Klein-Gordon equations - as well as any other Helmholtz-like equation - were recently shown to be associated with exact sets of ray-trajectories (coupled by a "Wave Potential" function encoded in their structure itself) describing any kind of wave-like features, such as diffraction and interference. This property suggests to view Wave Mechanics as a direct, causal and realistic, extension of Classical Mechanics, based on *exact trajectories and motion laws* of point-like particles "piloted" by de Broglie's matter waves and avoiding the probabilistic content and the wave-packets both of the standard Copenhagen interpretation and of Bohm's theory.

RÉSUMÉ - On a démontré récemment que les équations indépendantes du temps de Schroedinger et de Klein-Gordon, ainsi que toutes les autres équations d'Helmholtz, sont associées à des systèmes de trajectoires hamiltoniennes couplées par une function (le "potentiel d'onde") codée dans leur structure même, qui permettent de décrire tous le phénomènes ondulatoires, comme le diffraction et l'interférence. Cette propriété suggère d'envisager la Mécanique Ondulatoire comme une extension directe, causale et réaliste de la Mécanique Classique (basée sur les trajectoires exactes de particules ponctiformes pilotées par des ondes materielles monochromatiques) évitant à la fois le probabilisme et les paquets d'ondes de l'interpretation de Copenhagen et de la théorie de Bohm.

**1 - Introduction**

*"[La Mécanique Quantique], que je connais bien, puisque je l'ai longtemps enseignée, est très puissante et conduit à un très grand nombre de prévisions exactes, mais elle ne donne pas, à mon avis, une vue exacte et satisfaisante des phénomènes qu'elle étudie. Cela est un peu comparable au rôle joué naguère par la thermodynamique abstraite des principes qui permettait de prévoir exactement un gran nombre de phénomènes et était par suite d'une grande utilité, mais qui ne donnait pas une idée exacte de la réalité moléculaire dont le lois de la thermodynamique des principes ne donnaient que les conséquences statistiques".*

(Louis de Broglie, 1972 [**1**])

Any kind of monochromatic wave phenomena may be dealt with, as we shall see, in terms of an exact, *ray-based* kinematics, encoded in the structure itself of Helmholtz-like equations. The ray trajectories and motion laws turn out to be coupled by a dispersive "*Wave Potential*" function, which is responsible for *any*

---

[*] Corresponding author: adriano.orefice@unimi.it

*typically wave-like features such as diffraction and interference*, while its absence or omission confines the description to the geometrical optics approximation.
We extend this wave property, in the present paper, to the case of Wave Mechanics, thanks to the fact that both the *time-independent* Schrödinger and Klein-Gordon equations (associating monochromatic *de Broglie's matter waves* [**2,3**] to particles of assigned total energy) are themselves Helmholtz-like equations, allowing to formulate the Hamiltonian dynamics of point-like particles in terms of *exact trajectories and motion laws* under the coupling action of a suitable *Wave Potential*, in whose absence they reduce to the usual laws of classical dynamics. We make use of *relativistic* equations, in agreement both with de Broglie's firm belief that Wave Mechanics is an essentially relativistic theory [**4-7**] and with the first (unpublished) approach considered by Schrödinger, before his non-relativistic choice [**8, 9**].
As long as the association of exact ray-trajectories with any kind of monochromatic waves was not yet known, the intrinsically probabilistic interpretation of Wave Mechanics turned out to be the most plausible one: if indeed, according to de Broglie, the current interpretation "*ne donne pas une vue exacte et satisfaisante des phénomènes qu'elle étudie*", which is the hidden reality? The newfound general possibility of ray-trajectories [**10-12**] appears now to suggest that an "*exacte et satisfaisante*" reality *does* exist, and may be obtained from *the same Wave-Mechanical equations* viewed so far as unavoidably probabilistic. We show in Sect.2 how to obtain, both in Classical and Wave-Mechanical cases, consistent sets of exact kinematic and/or dynamic ray-based equations, and discuss in Sects.3 and 4 the proposal of viewing Wave Mechanics as a simple extension of Classical Mechanics.

**2- Dispelling commonplaces on wave trajectories**

 **2.1 -** We shall assume, in the following, both *wave monochromaticity* (strictly required by any such typically wave-like features as diffraction and interference) and *stationary media* (usually imposed by the experimental set-up). Although our considerations may be easily extended to most kinds of waves, we shall refer in this sub-Section, in order to fix ideas, to *classical electromagnetic waves* traveling according to a scalar wave equation of the simple form

$$\nabla^2 \psi - \frac{n^2}{c^2} \frac{\partial^2 \psi}{\partial t^2} = 0 \;, \qquad (1)$$

where $\psi(x,y,z,t)$ represents any component of the electric and/or magnetic field and $n(x,y,z)$ is the (time independent) refractive index of the medium.
By assuming

$$\psi = u(\vec{r},\omega) e^{-i\omega t} \;, \qquad (2)$$

with $\vec{r} \equiv (x,y,z)$, we get from eq.(1) the well-known Helmholtz equation [**13**]

$$\nabla^2 u + (n k_0)^2 u = 0 \qquad (3)$$

(where $k_0 \equiv \dfrac{2\pi}{\lambda_0} = \dfrac{\omega}{c}$), and look for solutions of the (quite general) form

$$u(\vec{r},\omega) = R(\vec{r},\omega)\, e^{\,i\,\varphi(\vec{r},\omega)}, \qquad (4)$$

with real $R(\vec{r},\omega)$ and $\varphi(\vec{r},\omega)$, which represent respectively, *without any probabilistic meaning*, the amplitude and phase of the monochromatic waves.

CONTRARY TO THE COMMONPLACE that a treatment in terms of *ray-trajectories* is only possible for a limited number of physical cases (such as reflection and refraction) ascribed to the so-called *geometrical optics approximations,* eq.(3) was shown in Refs.[**10-12**] to determine the *stationary* frame on which an *exact*, *ray-based* description is possible. By defining, in fact, the *wave-vector*

$$\vec{k} = \vec{\nabla}\,\varphi(\vec{r},\omega)\,, \qquad (5)$$

a set of *rays*, orthogonal to the phase surfaces $\varphi(\vec{r},\omega) = const$, turns out to travel, in *stationary* media, along *stationary* trajectories given by a simple Hamiltonian system of *kinematical* equations (see **Appendix**), under the basic action of a "Wave Potential" function

$$W(\vec{r},\omega) = -\,\frac{c}{2k_0}\frac{\nabla^2 R(\vec{r},\omega)}{R(\vec{r},\omega)}\,, \qquad (6)$$

inducing a mutual *perpendicular coupling* between the *monochromatic* ray-trajectories, *which is the one and only cause of wave-like features such as diffraction and interference*. An important consequence of this *perpendicularity* is the fact of leaving the intensity of the ray velocity unchanged, *confining the coupling action to a mere deflection*, while any possible variation of the speed amplitude is due to the refractive index of the medium. The limit of *geometrical optics* is reached when the space variation length $L$ of the wave amplitude $R(\vec{r},\omega)$ turns out to satisfy the condition $k_0 L \gg 1$. In this case the role of the Wave Potential is negligible, and the rays travel independently from one another under the only action of the refractive index, according to the "*eikonal equation*" [**13**]

$$k^2 \equiv (\vec{\nabla}\,\varphi)^2 \simeq (n\,k_0)^2\,. \qquad (7)$$

**2.2 -** Let us pass now (for simplicy sake, but with no subtantial loss of generality) to the *dynamics of single* (*chargeless and spinless*) *particles* with rest mass $m_0$ and assigned energy $E$, launched into a force field deriving from a stationary potential energy $V(\vec{r})$. Their *relativistic* behavior is described by the *time-independent* Hamilton-Jacobi equation [**14-17**]

$$[\vec{\nabla} S(\vec{r},E)]^2 = [\frac{E - V(\vec{r})}{c}]^2 - (m_0\,c)^2 \qquad (8)$$

where the basic property of the function $S(\vec{r},E)$ is that the particle momentum is given by

$$\vec{p} = \vec{\nabla} S(\vec{r}, E). \qquad (9)$$

In other words, the Hamilton-Jacobi surfaces $S(\vec{r}, E) = const$, perpendicular to the momentum of the moving particles, *"pilot" them, in Classical Mechanics, along a set of pre-fixed trajectories,* determining also their motion laws.

One of the main forward steps in modern physics, giving rise to Wave Mechanics, was performed by de Broglie's association of *mono-energetic* material particles [**2,3**] with suitable *monochromatic "matter waves"*, according to the correspondence

$$\vec{p}/\hbar \equiv \vec{\nabla} S(\vec{r}, E)/\hbar \rightarrow \vec{k} \equiv \vec{\nabla}\varphi. \qquad (10)$$

The Hamilton-Jacobi surfaces $S(\vec{r}, E) = const$ were assumed therefore as the phase-fronts of these *matter waves*, while maintaining their original role of "piloting" the particles - **just as in Classical Mechanics** - according to eq.(9).

The successive step was the assumption [**8, 9**] that these monochromatic matter waves satisfy a Helmholtz-like equation of the form (3), and that *the laws of Classical Mechanics* - represented here by eq.(8) - provide *the eikonal approximation* of this equation. By recalling, therefore, eqs. (7)-(10), one may perform the replacement

$$(n k_0)^2 \cong k^2 \rightarrow p^2/\hbar^2 \equiv (\vec{\nabla}\frac{S}{\hbar})^2 \equiv [\frac{E - V(\vec{r})}{\hbar c}]^2 - (\frac{m_0 c}{\hbar})^2 \qquad (11)$$

into eq.(3), reducing it to the *time-independent* Klein-Gordon equation

$$\nabla^2 u + [(\frac{E-V}{\hbar c})^2 - (\frac{m_0 c}{\hbar})^2] u = 0, \qquad (12)$$

holding for the de Broglie waves associated with particles of total energy $E$ moving in a stationary external potential $V(\vec{r})$. The physical existence of the waves predicted by de Broglie was very soon confirmed, as is well known, by the experiments performed by Davisson and Germer on electron diffraction by a crystalline nickel target [**18**].

CONTRARY TO THE COMMONPLACE that no *exact* particle trajectory may be defined in a Wave-Mechanical description, the same treatment providing the stationary ray-trajectories of the Helmholtz eq.(3) may now be applied to the particle trajectories associated with the *Helmholtz-like* eq.(12). Recalling eqs.(4) and (10), we assume therefore, in eq.(12),

$$u(\vec{r}, E) = R(\vec{r}, E) e^{i S(\vec{r}, E)/\hbar}, \qquad (13)$$

where the real functions $R(\vec{r}, E)$ and $S(\vec{r}, E)$ represent, respectively, without any probabilistic meaning, the amplitude and phase of de Broglie's mono-energetic matter waves, whose objective reality is established beyond any doubt by their observed properties of diffraction and interference. After the separation of real and imaginary parts, eq.(12) splits then into the system

$$\begin{cases} \vec{\nabla} \cdot (R^2 \vec{p}) \equiv \vec{\nabla} \cdot (R^2 \vec{\nabla}S) = 2\,R\,\vec{\nabla}R \cdot \vec{\nabla}S + R^2\,\vec{\nabla} \cdot \vec{\nabla}S = 0 & (14) \\ (\vec{\nabla}S)^2 - [\dfrac{E-V}{c}]^2 + (m_0 c)^2 = \hbar^2 \dfrac{\nabla^2 R(\vec{r},E)}{R(\vec{r},E)} & (15) \end{cases}$$

and the differentiation $\dfrac{\partial H}{\partial \vec{r}} \cdot d\vec{r} + \dfrac{\partial H}{\partial \vec{p}} \cdot d\vec{p} = 0$ of the relation

$$H(\vec{r},\vec{p},E) \equiv V(\vec{r}) + \sqrt{(pc)^2 + (m_0 c^2)^2 - \hbar^2 c^2 \dfrac{\nabla^2 R(\vec{r},E)}{R(\vec{r},E)}} = E \qquad (16)$$

obtained from eq.(15) is seen to be satisfied by the exact Hamiltonian set of dynamical equations

$$\begin{cases} \dfrac{d\vec{r}}{dt} = \dfrac{\partial H}{\partial \vec{p}} \equiv \dfrac{c^2 \vec{p}}{E - V(\vec{r})} & (17) \\ \dfrac{d\vec{p}}{dt} = -\dfrac{\partial H}{\partial \vec{r}} \equiv -\vec{\nabla}V(\vec{r}) - \dfrac{E}{E - V(\vec{r})} \vec{\nabla} Q(\vec{r},E) & (18) \end{cases}$$

with

$$Q(\vec{r},E) = -\dfrac{\hbar^2 c^2}{2E} \dfrac{\nabla^2 R(\vec{r},E)}{R(\vec{r},E)} \;, \qquad (19)$$

describing the particle motion along a set of *stationary trajectories*. It is interesting to observe that eq.(17) coincides with the "*guidance velocity*" proposed by de Broglie in his relativistic "*double solution theory*" [**4-7**], and that $\vec{v} \equiv \dfrac{d\vec{r}}{dt} \neq \vec{p}/m$, although $\vec{v} \equiv \dfrac{d\vec{r}}{dt}$ maintains itself parallel to the momentum $\vec{p}$. The function $Q(\vec{r},E)$, which we call once more, for simplicity sake, *"Wave Potential"*, has the same basic structure and coupling role of the function $W(\vec{r},\omega) = -\dfrac{c}{2k_0} \dfrac{\nabla^2 R(\vec{r},\omega)}{R(\vec{r},\omega)}$ of eq.(6): *it has therefore not so much a "quantum", as a "wave" origin,* entailed into quantum theory by de Broglie's matter waves. Just like the external potential $V(\vec{r})$, the Wave Potential $Q(\vec{r},E)$ is "encountered" by the particles along their motion $\vec{r}(E,t)$, and plays, once more, the basic role of *mutually coupling the trajectories* relevant to each mono-energetic matter wave. Once more, the *presence* of the Wave Potential is *the one and only cause of diffraction and/or interference of the waves*, and *its absence reduces the system* (17)-(18) *to the classical set of dynamical equations, which constitute therefore*, as expected, its *geometrical optics* approximation. Another interesting observation is that such phenomena as diffraction and interference *do not directly concern particles, but their (stationary) trajectories.* The overall number of traveling particles is quite indifferent, and may even be vanishingly small, so that, for instance, speaking of *self-diffraction* of a single particle is quite inappropriate. Each particle simply follows, according to the dynamical motion laws (17)-(18), the stationary trajectory, *pre-fixed from the very outset*, along which it's launched:

although, in the XVIII century, Maupertuis attributed this kind of behaviour to the "*wise action of the Supreme Being*", we limit ourselves to state that it is encoded in the structure itself of Helmholtz-like equations.

Eq.(14) plays the double role of *"closing"* the Hamiltonian system (17)-(18) by providing step by step, after the assignment of the wave amplitude distribution $R(\vec{r},E)$ over a launching surface, the necessary and sufficient condition for the determination of $R(\vec{r},E)$ over the next wave-front (thus allowing a consistent "closure" of the Hamiltonian system), and of <u>allowing</u> to show that the coupling "force" $\vec{\nabla}Q(\vec{r},E)$, of wave-like origin, is *perpendicular* to the particle momentum $\vec{p}$, so that that *no energy exchange is involved by its merely deflecting action*: any possible energy change may only be due to the external field $V(\vec{r})$. The assignment of the distribution $R(\vec{r},E)$ on the launching surface has the role, in its turn, of describing the essentials of the experimental set up. Let us finally notice that, in the particular case of *massless* particles (i.e. for $m_0 = 0$), the Klein-Gordon equation (12), by assuming the Planck relation

$$E = \hbar\omega, \qquad (20)$$

reduces to the form

$$\nabla^2 u + (n\omega/c)^2\, u = 0, \qquad (21)$$

with

$$n(\vec{r},E) = 1 - V(\vec{r})/E. \qquad (22)$$

Eq.(21) coincides with eq.(3), which may be therefore viewed as the time-independent Klein-Gordon equation holding for massless point-like particles in a stationary medium.

**2.3** - The same procedure applied in Sect.2.2 to obtain the stationary Klein-Gordon equation (12) was applied by Schrödinger [**8, 9**] to obtain his *non-relativistic* time-independent equation

$$\nabla^2 u(\vec{r},E) + \frac{2m}{\hbar^2}[E - V(\vec{r})]\, u(\vec{r},E) = 0 \qquad (23)$$

from the *non-relativistic* time-independent Hamilton-Jacobi equation

$$(\vec{\nabla}S)^2 = 2m[E - V(\vec{r})], \qquad (24)$$

and the association of eq.(23) with the *non-relativistic* dynamical system

$$\begin{cases} \dfrac{d\vec{r}}{dt} = \dfrac{\partial H}{\partial \vec{p}} \equiv \dfrac{\vec{p}}{m} & (25) \\[6pt] \dfrac{d\vec{p}}{dt} = -\dfrac{\partial H}{\partial \vec{r}} \equiv -\vec{\nabla}[V(\vec{r}) + Q(\vec{r},E)] & (26) \\[6pt] \vec{\nabla}\cdot(R^2\,\vec{p}) \equiv 0 & (27) \end{cases}$$

was shown in Refs.[**10-12**], in terms of the trajectory-coupling "Wave Potential"

$$Q(\vec{r},E) = -\frac{\hbar^2}{2m}\frac{\nabla^2 R(\vec{r},E)}{R(\vec{r},E)} \qquad (28)$$

and of the Hamiltonian function

$$H(\vec{r},\vec{p},E) = \frac{p^2}{2m} + V(\vec{r}) + Q(\vec{r},E) \qquad (29)$$

The time-independent Schrödinger equation (23) directly provides, in conclusion, the *exact, non-probabilistic point-particle* dynamical laws (25)-(27), reducing to the point-particle Hamiltonian/Newtonian description when the Wave Potential $Q(\vec{r},E)$ is neglected, i.e. in the limit of geometrical optics.

**2.4** - Many examples of numerical solution of the Hamiltonian particle dynamical system (25)-(27) in cases of diffraction and/or interference were given in Refs.[**10-12**], by assuming, for simplicity sake, the absence of external fields and a geometry allowing to limit the computation to the *(x,z)*-plane, for waves launched along the *z*-axis. The particle trajectories and the corresponding evolution both of de Broglie's wave intensity and of the Wave Potential were computed, with initial momentum components $p_x(t=0)=0$; $p_z(t=0)=p_0=2\pi\hbar/\lambda_0$, by means of a *symplectic* numerical integration method. We limit ourselves to present in the Figure the particle trajectories on the *(x,z)*-plane relevant to the diffraction of a *Gaussian* particle beam traveling along *z* and starting from a vertical slit centered at $x=z=0$ in the form $R(x;z=0) \div exp(-x^2/w_0^2)$, where the length $w_0$ is the so-called *waist radius* of the beam.

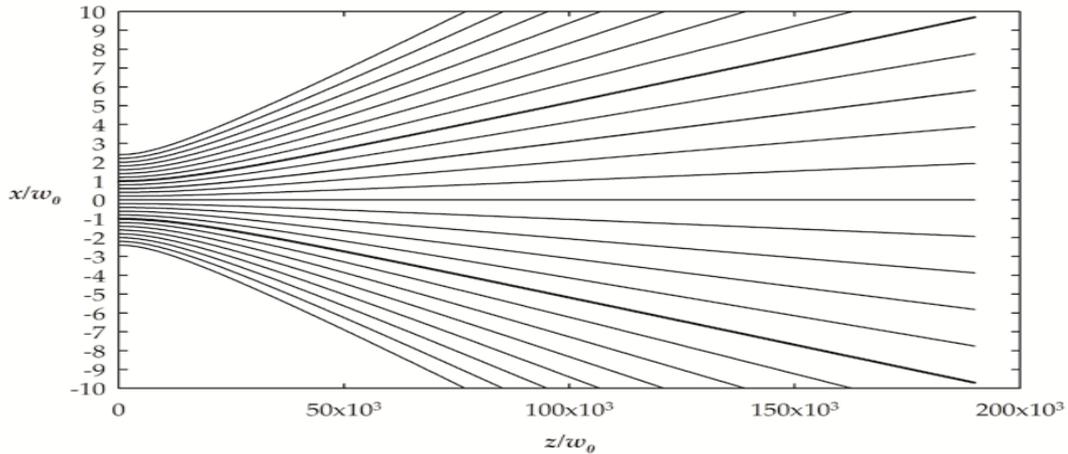

De Broglie's wave trajectories and waist lines on the symmetry *(x,z)*-plane for a *Gaussian* beam with waist $w_0$ and $\lambda_0/w_0 = 2\times10^{-4}$.

The two heavy lines are its *waist-lines*, given by the analytical relation

$$x(z) = \pm \sqrt{w_0^2 + \left(\frac{\lambda_0 \, z}{\pi \, w_0}\right)^2} \tag{30}$$

representing, in the so-called *paraxial approximation* [**19**], the trajectories starting (at $z = 0$) from the *waist positions* $x = \pm w_0$. The agreement between the *analytical* expression (30) and our *numerical* results provides, of course, an excellent test of our approach and interpretation. It was shown in Ref.[**12**] that the uncertainty relation $\Delta x \, \Delta p_x \geq \hbar$ *turns out to be violated close to the slit, but is asymptotically verified far enough from the slit,* thanks to the trajectory-coupling role of the Wave Potential $Q(\vec{r}, E)$.

### 3 - Born Rule vs Jaynes's "unscrambling prerequisite"

**3.1** - Let us now recall that, starting from eqs.(2), (20) and (23), one obtains [**14**] the *ordinary-looking* wave equation

$$\nabla^2 \psi = \frac{2m}{E^2}[E - V(\vec{r})] \frac{\partial^2 \psi}{\partial t^2}, \tag{31}$$

describing the propagation and dispersive character of *mono-energetic* de Broglie matter waves. By means, however, of the same eqs. (2), (20) and (23) one may also get the equation [**8, 9, 14, 15**]

$$\nabla^2 \psi - \frac{2m}{\hbar^2} V(\vec{r}) \, \psi = -\frac{2m}{\hbar^2} E \, \psi \equiv -\frac{2mi}{\hbar} \frac{E}{\hbar \omega} \frac{\partial \psi}{\partial t} = -\frac{2mi}{\hbar} \frac{\partial \psi}{\partial t}, \tag{32}$$

which is the usual form of the *time-dependent Schrödinger equation* for particles moving in a stationary potential field $V(\vec{r})$. Since eq.(32) is *not* a wave equation, any wave-like implication is due, in its case, to its connection with the *time-independent* Schrödinger equation (23), *from which it is obtained*. Eq.(23) admits indeed, as is well known, a (discrete or continuous, according to the boundary conditions) set of energy eigen-values and ortho-normal eigen-modes, which (referring for simplicity to the discrete case) we indicate respectively by $E_n$ and $u_n(\vec{r})$; and it's a standard procedure, making use of eqs.(2) and (20) and defining both the *eigen-frequencies* $\omega_n \equiv E_n/\hbar$ and the *eigen-functions*

$$\psi_n(\vec{r}, t) = u_n(\vec{r}) \, e^{-i \omega_n t} \equiv u_n(\vec{r}) \, e^{-i E_n t / \hbar}, \tag{33}$$

to verify that any linear superposition (with constant coefficients $c_n$) of the form

$$\psi(\vec{r}, t) = \sum_n c_n \, \psi_n(\vec{r}, t), \tag{34}$$

is a solution of eq.(32).
Since eqs.(31) and (32) hold *at same level of mathematical truism,* one may wonder what different roles they play in the treatment of de Broglie's waves.

At first glance, the time-dependent Schrödinger equation (32) merely describes the deterministic evolution of an *arbitrary* superposition of monochromatic waves $\psi_n(\vec{r},t)$, each one of which travels according to a wave equation of the form (31) (with $E = E_n$) along the Helmholtz trajectories determined by the relevant *time-independent* Schrödinger equation (23). Mainly because, however, of the energy-independence of eq.(32) and of the property of the $u_n(\vec{r})$ of constituting a complete ortho-normal basis, Born [20] proposed for the function (34) a role going much beyond that of a simple superposition: although eq.(32) is not - by itself - a wave equation, its solution (34) was assumed, under the name of "Wave Function", to represent *the most complete description of the physical state of a particle whose energy is not determined*: a vision giving to eq.(32) a dominant role both with respect to eq.(31) and to eq.(23). Even though "no generally accepted derivation has been given to date" [21], *this "Born Rule" aroused, together with Heisenberg's uncertainty relations, an intrinsically probabilistic conception of physical reality*, associating moreover to the *continuous and deterministic* evolution given by eq.(32) the *further Postulate* of a *discontinuous and probability-dominated* evolution process, after interaction with a measuring apparatus, in the form of a *collapse* (according to the probabilities $|c_n|^2$, in duly normalized form) into a single eigen-state. Because of the Born Rule a sharp distinction is made, therefore, between a superposition of *independent* mono-energetic waves $\psi_n(\vec{r},t)$ and their *inextricable coupling* (due to the assumption of $|\psi(\vec{r},t)|^2$ as a probability density) in a single "Wave Function" $\psi(\vec{r},t)$ evolving as a whole: a statistical mixture which becomes even more inextricable (and non-local) for a system of N particles, still assumed to be described by a single time-dependent Schrödinger equation and Wave Function.

The Born Rule provided a plausible interpretation of Schrödinger's equations (23) and (32) *as long as the possibility of associating exact ray-trajectories with any kind of Helmholtz-like equation was not yet known*. Such a newfound possibility suggests, however, a less striking interpretation of physical reality, limiting itself to view the function (34) (in duly normalized form) as a simple average taken over a superposition of *independent mono-energetic* waves $\psi_n(\vec{r},t)$, each one piloting point-like particles along exact Helmholtz trajectories pertaining to *their own* energy $E_n$, under the action of *their own* Wave Potential $Q(\vec{r},E_n)$. Being a merely mathematical assembling of either observational or hypothetical information (the set of coefficients $c_n$), this average is not bound, of course, to respect the locality properties of classical observables. Notice that, in any case, the property of any $\psi_n(\vec{r},t)$ of undergoing its own $Q(\vec{r},E_n)$ leads in general to the progressive spreading of the "wave-packet" $\psi(\vec{r},t)$.

Reminding that, according to E.T. Jaynes [22], "*our present quantum mechanical formalism is (...) an omelette that nobody has seen how to unscramble, and whose unscrambling is a prerequisite for any future advance in basic physical theory*", the mono-energetic de Broglie waves - when considered as *independent* from one another (i.e. not *scrambled* together) - allow a non-probabilistic description in terms of exact point-particle trajectories, providing a straightforward Wave-Mechanical extension of Classical Dynamics.

**3.2** - It's worthwhile reminding that, although the *time-dependent* Schrödinger equation (32) is a simple consequence of the *time-independent* equation (23), its "stronger" version

$$\nabla^2 \psi - \frac{2m}{\hbar^2} V(\vec{r},t)\,\psi = -\frac{2mi}{\hbar}\frac{\partial \psi}{\partial t} \quad , \tag{35}$$

containing a time-dependent external potential $V(\vec{r},t)$, may only be considered as a *Postulate*, and is often assumed, indeed, as a First Principle at the very beginning of standard textbooks.

Although it cannot be obtained from de Broglie's basic assumption (10), it was accepted by de Broglie himself [**5**] with these words: "*The form of* [eq.(32)] *allows us to go beyond single monochromatic waves and to consider superpositions of such waves. In addition, it suggests the way to extend the new Mechanics to the case of fields varying with time. Indeed, since it permits us to go beyond monochromatic waves, time no longer plays a special part, and* **it is then natural to admit** *that the form of the equation must be preserved when V depends on time as the general form of the equation of propagation of ψ waves in the non-relativistic Wave Mechanics of a single particle*". We limit ourselves to observe that eq.(35) cannot lose, in its induction from eq.(32), its statistical character.

**3.3** - Let us come, finally, to the case of the "*quantum trajectories*" of Bohm's theory **[23-34],** and to their possible connection with the present analysis, aiming to a simple extension of Classical Mechanics in terms of *exact* particle trajectories. According to Ref.[**31**] "*Born had an absolutely correct (...) intuition about the meaning of the wave function, namely that it guides the particles and it determines the probability of particle positions. (...) Born is close to Bohmian mechanics*". In Bohm's approach, indeed, a replacement of the form

$$\psi(\vec{r},t) = R(\vec{r},t)\,e^{\,i\,S(\vec{r},t)/\hbar} \tag{36}$$

is performed into eq.(35), splitting it [**23**], after separation of real and imaginary parts, into the coupled system

$$\begin{cases} \dfrac{\partial P}{\partial t} + \vec{\nabla}\cdot(P\dfrac{\vec{\nabla} S}{m}) = 0 & (37) \\[1em] \dfrac{\partial S}{\partial t} + \dfrac{(\vec{\nabla} S)^2}{2m} + V(\vec{r},t) - \dfrac{\hbar^2}{2m}\dfrac{\nabla^2 R}{R} = 0 & (38) \end{cases}$$

where, in agreement with Born's vision, the function $P(\vec{r},t) \equiv R^2(\vec{r},t)$ is assumed to represent, in Bohm's words, "*the probability density for particles belonging to a statistical ensemble*", and eq.(37) is viewed as a fluid-like continuity equation for such a probability density. Bohm's replacement (36) (shaped on eq.(13), i.e on de Broglie's mono-energetic pilot-waves, *whose objective reality is well-established by diffraction and interference experiments*) depicts $\psi(\vec{r},t)$ as a single physical wave (the Born Wave Function), *hopefully sharing the same objective reality*. Bohm's approach is indeed a strong attempt to dress with plausibility the Born Rule by presenting the function $\psi(\vec{r},t)$ as a *generalized* pilot-wave, although it isn't even the solution of a wave equation. Eq.(38) is viewed, in this spirit, as *analogous* to a time-dependent Hamilton-Jacobi equation, containing however, in addition to the external potential $V(\vec{r},t)$, a "Quantum Potential"

$$Q_B(\vec{r},t) = -\frac{\hbar^2}{2m}\frac{\nabla^2 R(\vec{r},t)}{R(\vec{r},t)} \qquad (39)$$

coupling together, when $V=V(\vec{r})$, the full set of eigen-functions $\psi_n(\vec{r},t)$. Notice that while the (formally similar) mono-energetic *Wave Potential* $Q(\vec{r},E)$ of eq.(28) couples *perpendicularly* (and therefore without any wave-particle energy exchange) the relevant mono-energetic trajectories, the *Quantum Potential*, because of its composite structure, doesn't appear to play such a "soft piloting" role along Bohm's "quantum trajectories", unless it simply corresponds to a superposition of *independent* mono-energetic waves, each one of which "pilots" particles without any energy exchange.

The presence (or absence) of energy exchanges due to wave-mechanical interactions (and supporting either an energy-less *Wave Potential* superposition or an energy-exchanging *Quantum Potential* "scrambling") could be ascertained by means of experimental tests in nano-technology, as, for instance, in the case of the ion beams [**35**] employed to prepare the single parts of the compact integrated circuits (microchips), containing billions of electronic components per square centimeter, nowadays employed in virtually all electronic equipments.

A so-called "*guidance equation*", originally proposed by de Broglie himself [**4-7**] in a quite different context, is finally assumed, in Bohm's theory, in the form

$$\frac{d\vec{r}}{dt} = \vec{v}(\vec{r},t) \equiv \vec{\nabla}S(\vec{r},t)/m \equiv \frac{\hbar}{m}Im(\frac{\vec{\nabla}\psi}{\psi}) \,, \qquad (40)$$

making use of a well-known expression of quantum *probability flow* [**14**, **15**], and coupling together, just like $Q_B(\vec{r},t)$, the full set of eigen-functions $\psi_n(\vec{r},t)$.

Bohm's theory makes use moreover, in the case of a system of N particles, of N *guidance equations* of the form (40), coupled - as in the Copenhagen description - by a single Wave Function. Eq.(40) is then time-integrated in parallel with the solution of the relevant time-dependent Schrödinger equation, bypassing the direct use of the Quantum Potential $Q_B(\vec{r},t)$. The Wave Function $\psi(\vec{r},t)$ is therefore associated step-by-step with a set of "*quantum trajectories*" $\vec{r}(t)$, representing *"the flux lines along which the probability is transported"* [**31**] and providing a continuous process which culminates - without resorting to a Wave Function collapse - at the final state. In Bohm's words [**27**], "*the guidance conditions and the quantum potential depend on the state of the whole system in a way that cannot be expressed as a preassigned interaction between its parts. As a result there can arise a new feature of objective wholeness. This is not only a consequence of the non-local interactions but even more it follows from the fact* *that the entire system of particles is organised by a common "pool of active information" which does not belong to the set of particles but which, from the very outset, belongs to the whole***".**

**3.4 -** Despite the evident connection between the (simple) <u>Helmholtz coupling</u> among mono-energetic ray-trajectories (which *does* " belong to the whole ", from the very outset, for any stationary wave, and is well-established by diffraction and interference experiments), and the (inextricable) <u>Bohm/Born coupling</u> among different parts of a quantum system, we conclude that Bohm's approach (through the Born Rule, the probabilistic definition of $P(\vec{r},t)$ and the "guidance condition") is far from the spirit of the present paper, whose aim is to suggest *an exact, non-probabilistic Wave Mechanics, running as close as possible to Classical Mechanics*. The Bohmian theory declaredly provides, on the contrary, an equivalent version of the (intrinsically probabilistic) Copenhagen vision, including most "quantum paradoxes" [**36**].

We are *not proposing* an equivalent route (in the sense, for instance, of the Newtonian/Hamiltonian equivalence in Classical Mechanics), or a different level of approximation: *we are proposing a different model of physical reality*. We synthetize the two approaches in the following ("Exact" vs. "Probabilistic") Tables, comparing the respective (non-relativistic) sets of equations for the motion of single particles in an external potential $V(\vec{r})$:

| TAB.I<br><br>EXACT (POINT-PARTICLE) DESCRIPTION | TAB.II<br><br>PROBABILISTIC (WAVE-PACKET) DESCRIPTION |
|---|---|
| $$\frac{d\vec{r}}{dt} = \frac{\vec{p}}{m}$$ $$\frac{d\vec{p}}{dt} = -\vec{\nabla}[\ V(\vec{r}) - \frac{\hbar^2}{2m}\frac{\nabla^2 R(\vec{r},E)}{R(\vec{r},E)}\ ]$$ $$\vec{\nabla}\cdot(R^2\ \vec{p}\ ) = 0$$ | $$i\hbar\ \frac{\partial\psi}{\partial t} = -\frac{\hbar^2}{2m}\nabla^2\psi + V(\vec{r})\ \psi$$ $$\frac{d\vec{r}}{dt} = \frac{\hbar}{m}\ \text{Im}\ \frac{\vec{\nabla}\psi(\vec{r},t)}{\psi(\vec{r},t)}$$ |

The equations of TAB.I are *encoded in Schrödinger's (Helmholtz-like) time-independent equation*, and provide the *exact trajectories of a point-particle* of assigned energy *E*, guided by a de Broglie matter wave of amplitude *R* whose objective reality is shown by its diffractive properties.

The equations of TAB.II provide the *probability flux-lines* of a *wave packet* (resulting from the entire set of stationary eigen-solutions) built up step-by-step as a whole by *the simultaneous solution of Schrödinger's time-dependent equation*, adding therefore further details to the statistical formalism of Quantum Mechanics. No *a priori* energy value may be assigned, since the wave packet represents the maximum possible information about the intrinsically uncertain reality of the particle.

## 4 - Discussion and conclusions

We summarize the content of the present paper as follows:

- An exact, ray-based kinematics is encoded in the structure itself of Helmholtz-like equations - employed all over Classical Mechanics for any kind of monochromatic waves - without any need of probabilistic assumptions.
- Since both the time-independent Schrödinger and Klein-Gordon equations belong to the family of Helmholtz-like equations, the dynamics of particles associated with mono-energetic de Broglie's waves may be expressed in terms of ray-based, non-probabilistic motion laws, generalizing (through the role of the Wave Potential) the equations of Classical Dynamics, and preserving the same exact and objective realism.
- Although we limited ourselves for simplicity sake (but without a substantial loss of generality) to consider the motion of single particles, we deem therefore that the intrinsically probabilistic nature of physical reality claimed by the current Quantum Mechanical vision is by no means a universal property, since it may be excluded in the most simple cases. One can only maintain that the *time-dependent* Schrödinger equation, both in the standard and in the Bohmian quantum approaches, provides a statistical, *coarse-grained* description of a *fine-grained* underlying reality. A description which may represent, as in the case of classical Thermodynamics [**1**], the best way to deal with a wide class of problems, but is not concerned with an exact and exhausting picture of that reality.

A discussion concerning the compatibility with the Heisenberg relations is here beyond our purposes. Their connection with the Wave Potential, on the one hand, is sketched in Ref.[**12**], and their *disturbance interpretation*, on the other hand, is nowadays quite generally recognized [**37, 38**] to be untenable. One could distinguish, of course, between the notions of *state preparation* of the experiment and of *state-disturbing* measurement [**39**]; but we avoid any distinction and opt, in conclusion, for a *vision of Wave Mechanics allowing to satisfy Jaynes's "unscrambling prerequisite" by means of a straightforward, ray-based extension of Classical Mechanics.*

# APPENDIX

After the use of eq.(4) and the separation of real and imaginary parts, the Helmholtz eq.(3) splits into the system

$$\begin{cases} \vec{\nabla} \cdot (R^2 \vec{\nabla} \varphi) \equiv 2R\vec{\nabla}R \cdot \vec{\nabla}\varphi + R^2 \vec{\nabla} \cdot \vec{\nabla}\varphi = 0 & \text{(A1)} \\ D(\vec{r},\vec{k},\omega) \equiv \dfrac{c}{2k_0}[k^2 - (nk_0)^2] + W(\vec{r},\omega) = 0 & \text{(A2)} \end{cases}$$

where

$$W(\vec{r},\omega) = -\frac{c}{2k_0}\frac{\nabla^2 R(\vec{r},\omega)}{R(\vec{r},\omega)} \ , \tag{A3}$$

and the differentiation $\dfrac{\partial D}{\partial \vec{r}} \cdot d\vec{r} + \dfrac{\partial D}{\partial \vec{k}} \cdot d\vec{k} = 0$ of eq.(A2) is satisfied by the kinematic Hamiltonian system

$$\begin{cases} \dfrac{d\vec{r}}{dt} = \dfrac{\partial D}{\partial \vec{k}} \equiv \dfrac{c\vec{k}}{k_0} & \text{(A4)} \\ \dfrac{d\vec{k}}{dt} = -\dfrac{\partial D}{\partial \vec{r}} \equiv \vec{\nabla}[\dfrac{ck_0}{2}n^2(\vec{r},\omega) - W(\vec{r},\omega)] & \text{(A5)} \end{cases}$$

associating with the Helmholtz equation an *exact stationary set of trajectories* along which the *monochromatic* rays (each one characterized by its launching position and wave vector) are driven. Since no new trajectory may arise in the space region spanned by the wave trajectories (so that $\vec{\nabla} \cdot \vec{\nabla}\varphi = 0$), eq.(A1) tells us that $\vec{\nabla}R \cdot \vec{\nabla}\varphi = 0$: the amplitude $R(\vec{r},\omega)$ of the monochromatic wave - together with its derivatives and functions, including $W(\vec{r},\omega)$ - is therefore distributed over the relevant wave-front, normal to $\vec{k} \equiv \vec{\nabla}\varphi(\vec{r},\omega)$, and the coupling term $\vec{\nabla}W(\vec{r},\omega)$ acts *perpendicularly* to the relevant ray trajectories. Eq.(A1) provides moreover step by step, after the assignment of the wave amplitude distribution $R(\vec{r},\omega)$ over the launching surface, the necessary and sufficient condition for the determination of $R(\vec{r},\omega)$ over the next wave-front, thus allowing a consistent *"closure"* of the Hamiltonian system. When, in particular, the space variation length $L$ of the wave amplitude $R(\vec{r},\omega)$ turns out to satisfy the condition $k_0 L \gg 1$, eq.(A2) reduces to the well-known *eikonal equation* [13]

$$k^2 \equiv (\vec{\nabla}\varphi)^2 \simeq (nk_0)^2 . \tag{A6}$$

In this *geometrical optics approximation* the coupling role of the Wave Potential is neglected, and the rays travel independently from one another under the only action of the refractive index.